\documentclass[manuscript,screen]{acmart}
\usepackage{graphicx, comment}
\usepackage{bm}
\usepackage{color}
\usepackage{epstopdf}
\usepackage{epsfig}
\usepackage{verbatim}
\usepackage{lineno}
\usepackage{subcaption}
\usepackage[thicklines]{cancel}
\usepackage{url}
\usepackage{soul}
\usepackage{algorithm, algorithmic}
\usepackage[braket, qm]{qcircuit}
\pdfmapfile{+sansmathaccent.map}
\usepackage{tikz}
\usepackage[edges]{forest}
\usetikzlibrary{arrows.meta,shadows.blur}
\colorlet{myyellow}{yellow!50}
\usepackage{float}
\usepackage{dblfloatfix}
\usepackage{graphicx,float}

\usetikzlibrary{shapes,arrows}
\usepackage{amsmath,bm,times}
\usepackage{pgfplots}
\usepackage{pgfplotstable}
\usepackage{subcaption}
\usepackage[export]{adjustbox}
\usepackage{bera}
\usepackage{listings,mdframed}
\usepackage{xcolor}
\usepackage{pythonhighlight}
\definecolor{comment-text-color}{rgb}{0,0.8,0.6}
\lstdefinelanguage{json}{
    basicstyle=\normalfont\ttfamily\footnotesize,
    numbers=left,
    numberstyle=\scriptsize,
    stepnumber=1,
    numbersep=8pt,
    showstringspaces=false,
    breaklines=true,
    frame=lines,
    backgroundcolor=\color{background},
    literate=
     *{0}{{{\color{numb}0}}}{1}
      {1}{{{\color{numb}1}}}{1}
      {2}{{{\color{numb}2}}}{1}
      {3}{{{\color{numb}3}}}{1}
      {4}{{{\color{numb}4}}}{1}
      {5}{{{\color{numb}5}}}{1}
      {6}{{{\color{numb}6}}}{1}
      {7}{{{\color{numb}7}}}{1}
      {8}{{{\color{numb}8}}}{1}
      {9}{{{\color{numb}9}}}{1}
      {:}{{{\color{punct}{:}}}}{1}
      {,}{{{\color{punct}{,}}}}{1}
      {\{}{{{\color{delim}{\{}}}}{1}
      {\}}{{{\color{delim}{\}}}}}{1}
      {[}{{{\color{delim}{[}}}}{1}
      {]}{{{\color{delim}{]}}}}{1},
}

\usepackage{xcolor}
\usepackage{listings}
\usepackage{enumerate}
\usepackage{wrapfig}
\usepackage{siunitx}
\usepackage{booktabs}
\usepackage{hyperref}
\hypersetup{
     colorlinks   = true,
     citecolor    = blue
}

\definecolor{red}{rgb}{1,0.,0}

\usepackage[compatibility=false]{caption}
\usepackage{tcolorbox}
\newlength\myindent
\newcommand\bindent{%
  \begingroup
  \setlength{\itemindent}{\myindent}
  \addtolength{\algorithmicindent}{\myindent}
}
\newcommand\eindent{\endgroup}
\begin{document}

\title{Extending Python for Quantum-Classical Computing via Quantum Just-in-Time Compilation}

\thanks{This manuscript has been authored by UT-Battelle, LLC under Contract No. DE-AC05-00OR22725 with the U.S. Department of Energy. The United States Government retains and the publisher, by accepting the article for publication, acknowledges that the United States Government retains a non-exclusive, paid-up, irrevocable, world-wide license to publish or reproduce the published form of this manuscript, or allow others to do so, for United States Government purposes. The Department of Energy will provide public access to these results of federally sponsored research in accordance with the DOE Public Access Plan. (http://energy.gov/downloads/doe-public-access-plan).}

\begin{abstract}
Python is a popular programming language known for its flexibility, usability, readability, and focus on developer productivity. The quantum software community has adopted Python on a number of large-scale efforts due to these characteristics, as well as the remote nature of near-term quantum processors. The use of Python has enabled quick prototyping for quantum code that directly benefits pertinent research and development efforts in quantum scientific computing. However, this rapid prototyping ability comes at the cost of future performant integration for tightly-coupled CPU-QPU architectures with fast-feedback. Here we present a language extension to Python that enables heterogeneous quantum-classical computing via a robust C\texttt{++} infrastructure for quantum just-in-time (QJIT) compilation. Our work builds off the QCOR C\texttt{++} language extension and compiler infrastructure to enable a single-source, quantum hardware-agnostic approach to quantum-classical computing that retains the performance required for tightly coupled CPU-QPU compute models. We detail this Pythonic extension, its programming model and underlying software architecture, and provide a robust set of examples to demonstrate the utility of our approach. 
\end{abstract}

\author{Thien Nguyen}
\affiliation{Computer Science and Mathematics Division,\ Oak\ Ridge\ National\ Laboratory,\ Oak\ Ridge,\ TN,\ 37831,\ USA}

\author{Alexander J.\ McCaskey}
\email{mccaskeyaj@ornl.gov}
\affiliation{Quantum Computing Institute,\ Oak\ Ridge\ National\ Laboratory,\
  Oak\ Ridge,\ TN,\ 37831,\ USA}
\affiliation{Computer Science and Mathematics Division,\ Oak\ Ridge\ National\ Laboratory,\ Oak\ Ridge,\ TN,\ 37831,\ USA}

\maketitle

\section{Introduction}
The availability of noisy quantum hardware has necessitated the concurrent advance of corresponding software and tooling for enabling remote programming, compilation, and execution. Vendors have put forward concrete, noisy implementations of superconducting, ion trap, and photonic quantum computers, and most have provided corresponding software frameworks for low-level circuit construction and remote execution over HTTP REST APIs. This concurrent hardware-software implementation strategy has resulted in a number of high-level scientific computing demonstrations pertinent to fields like nuclear physics, quantum chemistry, and machine learning \cite{Dumitrescu2018,McCaskey2019,PhysRevA.98.032331,hamilton}. The software frameworks put forward over the past few years have primarily leveraged Python to provide a high-level language approach to quantum-classical programming~\cite{Qiskit, pyquil, projectQ, cirq, Strawberry_Fields}. The decision to use Python makes sense for near-term quantum-classical computing tasks for a number of reasons: (1) the remote nature of the CPU-QPU interaction removes any possibility of performance (therefore a less-performant language like Python suffices), (2) the wide availability of libraries for common tasks like network programming, (3) the experimental nature of near-term quantum processors necessitates the quick-prototyping capability provided by scripting, and (4) the grass-roots development of tutorials and teaching materials for quantum programming puts a strong focus on easy-to-learn scripting languages commonly taught in University computer science programs.

A consequence of the wealth of development activities coming from these separate vendor framework efforts has been the lack of a unified interface, a true integration platform, for near-term quantum computation at the Pythonic language level. A number of frameworks have moved toward an architectural model that enables backend injection via the development of pertinent subclasses, however, each framework is primarily focused on providing strong support for the hardware backend it was originally intended for. Programmers, therefore, are forced to switch frameworks - learn new data structures and models - every time they want to program a different quantum computer, if they desire to get the most out of the feature set provided by the framework. This has a large effect on algorithmic and benchmarking research activities that require the ability to quickly prototype algorithms and applications and compare execution across a variety of near-term QPUs. 

Here we present a programming model that builds upon the \texttt{qcor} C\texttt{++} language extension~\cite{mintz2019qcor, qcor-paper} to provide a unified interface and integration platform for quantum-classical computing at the Python language level. Specifically, we provide a language extension to Python that enables a just-in-time (JIT), retargetable compiler for quantum-classical Python scripts that wholly delegates to a performant C\texttt{++} infrastructure. Our approach allows programmers to write quantum-classical codes at a high-level in Python in a way that supports quick prototyping, library integration, and dynamic typing in a write-once, run-anywhere manner. Our approach is truly hardware-agnostic, with backend support from IBM, Rigetti, Honeywell, and others, as well as Summit-scale quantum circuit simulation technologies. 

This paper is structured as follows: first we provide relevant background on the \texttt{qcor} C\texttt{++} language extension and compiler infrastructure. We then detail how we extend that infrastructure for Pythonic quantum kernel parsing and compilation in C\texttt{++}. Next we provide a detailed exposition of the \texttt{qcor} just-in-time compilation infrastructure, enabling the compilation and execution of quantum kernel strings at runtime. Finally, with this background in place, we detail how we provide our Pythonic language extension (quantum kernels in Python) via delegation to this C\texttt{++} JIT infrastructure. We end with a demonstration detailing pertinent examples of the utility of this work. 

\section{QCOR}
Recently, a language extension specification \cite{mintz2019qcor} has been proposed for heterogeneous quantum-classical computing that seeks to enable a single-source, accelerated-node programming model by leveraging language-native functions to express quantum code (kernels) intended for compilation and execution on a quantum co-processor. This specification, QCOR, has recently been implemented as an extension to C\texttt{++} via a compiler infrastructure called \texttt{qcor} \cite{qcor-paper}. The \texttt{qcor} compiler enables the expression of quantum kernels as standalone, annotated functions in C\texttt{++} and compiles these hybrid quantum-classical source files in a manner that enables quantum backend retargetability. \texttt{qcor} achieves this functionality by extending key plugin interfaces in Clang that enable quantum kernel language parsing and syntax handling at compile-time (after preprocessing, before abstract-syntax-tree generation). The Clang \texttt{SyntaxHandler} extension point \cite{clangsh} provides implementors with an opportunity to analyze invalid domain specific language code and map it to appropriate, valid C\texttt{++} API calls. \texttt{qcor} puts forward a \texttt{SyntaxHandler} that maps invalid quantum code to valid API calls that target the XACC quantum-classical programming framework \cite{mccaskey2020xacc}. In this way, the \texttt{qcor} compiler represents the integration of classical compiler software technologies with the XACC programming model and concrete implementation for gate-model quantum-classical computing. XACC is leveraged by  \texttt{qcor} for its quantum intermediate representation (IR), IR transformation infrastructure, backend quantum co-processor extensibility, and language parsing capabilities. 

Moreover, the \texttt{qcor} \texttt{SyntaxHandler} implementation is further decomposed into a unique \texttt{TokenHandler} interface which is implemented for new quantum domain specific languages, thereby enabling extensible quantum language expression and utility within quantum kernel function bodies. The goal of a \texttt{TokenHandler} sub-type is to analyze incoming observed Clang \texttt{Tokens} and provide valid C\texttt{++} code as a replacement that affects construction of the XACC IR and its execution on the desired backend quantum co-processor. \texttt{qcor} currently has \texttt{TokenCollector} implementations that enable kernel programming using the XASM and OpenQASM 2.0 assembly dialects, as well as a unitary matrix decomposition language for high-level quantum program expression. As an extension point to the \texttt{qcor} compiler platform, developers are free to enable quantum code expression in a way that best suits the situation at hand. For this work - extending Python with support for quantum kernel expression - we seek a \texttt{TokenCollector} implementation that will parse Pythonic quantum assembly expressions with support for typical Pythonic control flow constructs like \texttt{for} and \texttt{if}. 

The following sections will detail how we leverage the \texttt{TokenCollector} to enable quantum kernel programming in C\texttt{++} using a Pythonic version of the XASM dialect. With this in place, we detail how to leverage these expressions via just-in-time compilation of \texttt{qcor} quantum kernels (QJIT). Finally, we show how this QJIT infrastructure is leveraged from Python to enable high-level Pythonic quantum kernel expression and utility via standard Python function decorators.

\subsection{New \texttt{qcor} C\texttt{++} Language Features}
\label{sec:qcor_update}
Before diving into the specifics of our Python language extension for quantum computing based on \texttt{qcor}, it is pertinent to mention new features that have made it into the C\texttt{++} language extension since the publication of \cite{qcor-paper}. We note these to raise awareness of them, but also because a goal of the Python extension is to support all features that the C\texttt{++} extension supports. The added features described in this section are also provided by the Pythonic extension described by this work.

\subsubsection{Compute - Action - Uncompute Programming Pattern}
\label{sec:qcor_compute_action}
A common pattern in quantum computing is the $U-V-U^\dagger$ sequence~\cite{H_ner_2018}, with unitary matrices $U$ and $V$. It is often referred to as the \emph{compute-action-uncompute} pattern due to the cancellation effect of $U^\dagger$. For example, quantum circuits implementing reversible logic~\cite{5391327} usually have this structure. A recent enhancement to the \texttt{qcor} compiler frontend and its runtime allows users to capture this specific pattern by introducing the \textbf{\texttt{compute}} and \textbf{\texttt{action}} keywords. The language expects a brace-enclosed sub-kernel, e.g., quantum gates, after each of these two keywords. The uncompute circuit, i.e., the adjoint of the compute gate sequence, is appended automatically after the action block by the compiler. Once the compiler detects this pattern, it marks the compute sub-circuit segment by injecting calls to the internal \texttt{qcor} runtime functions at the beginning and the end of the sub-circuit. Thanks to these markers, the \texttt{qcor} runtime will be able to collect the gate sequence representing the compute $U$ matrix and then append its conjugate after the action segment. 

In addition to improving the expressibility of the language, this \texttt{compute}-\texttt{action} construct also benefits controlled circuit generation. Due to the cancellation effect of the $U-U^\dagger$ pair, we do not need to add control qubits to these two segments but just the middle action $V$ segment. Therefore, the resulting controlled circuit is potentially much shorter than a naive gate-by-gate transformation without this high-level structural meta information.

\subsubsection{Callable Quantum Kernel Arguments and Lambda Expressions}
\label{sec:qcor_QuantumSignature}
A major component of the \texttt{qcor} compiler platform is its standard library implementation providing commonly used subroutines, such as Grover's algorithm, quantum Fourier transforms, error correction procedures, etc. To develop generic, library-oriented kernels, it is imperative to be able to pass quantum kernels as arguments to other kernels. To address this requirement, we introduce a new core concept into \texttt{qcor} --- the \texttt{KernelSignature}, which is very similar to the \texttt{QuantumKernel} class~\cite{qcor-paper} but intended for wrapping quantum kernels as type-safe callable objects. \texttt{KernelSignature} objects are implicitly constructible from \texttt{qcor} kernel functions (see Fig. \ref{fig:lambda_sample} \texttt{oracle\_as\_function} and its usage with \texttt{phase\_estimation}). In other words, users can directly pass a kernel by its name to an argument of type \texttt{KernelSignature}.

\begin{wrapfigure}{r}{.5\textwidth}
  \lstset {language=C++}
  \begin{lstlisting}
__qpu__ void phase_estimation(qreg q, 
            KernelSignature<qubit> oracle) {
... // Omitted
}

__qpu__ void oracle_as_function(qubit q) { 
  T(q); 
}

int main(int argc, char **argv) {
  auto q = qalloc(4);
  // Provide oracle as function
  phase_estimation(q, oracle_as_function);
  
  // Provide oracle as a lambda
  int capture_int = 5;
  auto oracle = 
        qpu_lambda([](qubit q) { 
          print(capture_int); 
          T(q); 
        }, 
        capture_int);
  phase_estimation(q, oracle);
}
\end{lstlisting}
\caption{Here we show how one can write quantum kernels parameterized on other quantum kernels via the \texttt{KernelSignature<T...>} type. We also leverage this for passing quantum lambda expresssions.}
\label{fig:lambda_sample}
\end{wrapfigure}
Another advanced yet ubiquitous feature of many high-level programming languages is the lambda expression, which enables anonymous functions with the ability to capture variables from the surrounding scope in addition to specified function arguments. Supporting quantum lambda kernels in \texttt{qcor} enables a more concise syntax for a variety of use cases, as shown in Fig.~\ref{fig:lambda_sample}. In this example, the \texttt{oracle} kernel is defined in place by using the newly-added \texttt{qpu\_lambda} feature and passed to a generic quantum phase estimation kernel which expects an \texttt{oracle} argument of type \texttt{KernelSignature<qubit>}. We want to note that lambda capture is not required in this case but it is a very useful feature to bind additional variables from the local scope to an otherwise fixed functional signature.  

Since the \texttt{SyntaxHandler} extension of Clang is designed for free-standing functions, we need a different set of machinery to handle quantum lambda expression, namely macro pre-processing and LLVM Just-In-Time (JIT) engine. Concretely speaking, the \texttt{qpu\_lambda} macro turns the quantum lambda body into a source string which will be JIT-compiled, with proper syntax handling, and returns an internal lambda kernel class instance supporting call-like invocation. By-value or by-reference capture variables, if any, can be appended to the functional signature of the underlying rewritten kernel function using their exact names. This allows us to emulate variable capturing in a free-standing, capture-less \texttt{qcor} function. The \texttt{qcor} lambda object keeps a handle to the JIT-compiled function along with copies of any captured variables, and is therefore able to provide the same invocation functionality as one would see with conventional C\texttt{++} lambdas. As depicted in Fig.~\ref{fig:lambda_sample}, \texttt{qpu\_lambda} is compatible with \texttt{KernelSignature} argument type similar to global scope \texttt{\_\_qpu\_\_} kernel functions.

\section{Pythonic Kernel Processing in C\texttt{++}}
\begin{wrapfigure}{r}{.6\textwidth}
\centering  
\includegraphics[width=.6\textwidth]{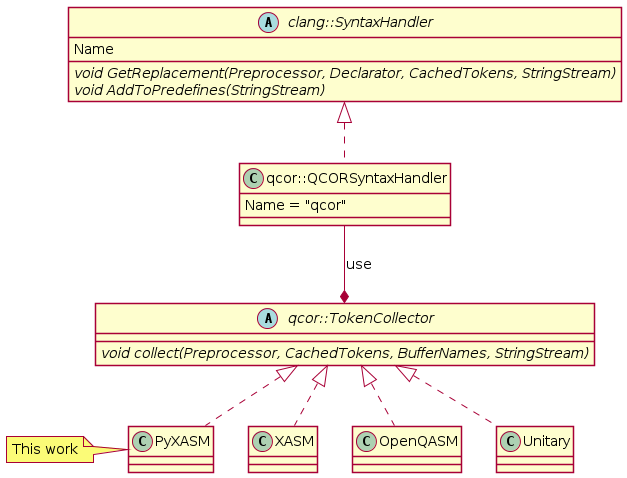}
\caption{\texttt{qcor} provides a single-source C\texttt{++} programming via the Clang \texttt{SyntaxHandler} extension.}
\label{fig:qcor_layers}
\end{wrapfigure}

The first step in creating our Python language extension is to extend the \texttt{qcor} kernel processing infrastructure with support for a Pythonic quantum assembly dialect. Specifically, we will attempt to provide the same functionality inherent to the C\texttt{++} XASM dialect (simple quantum instructions and C\texttt{++} control flow, variable assignment, class instance method invocation, etc.), but with a syntax that will ultimately be valid with respect to the Python interpreter. To do this, we implement the \texttt{TokenCollector} interface and override the \texttt{collect()} method, see Fig.~\ref{fig:qcor_layers}, to analyze incoming Pythonic quantum assembly tokens (\texttt{clang::Token}) and provide a re-written source string for use by the Clang \texttt{SyntaxHandler} infrastructure. Ultimately the goal of this \texttt{TokenCollector} is to map incoming Pythonic quantum code into corresponding C\texttt{++}. 

There are a few key challenges in performing this mapping. The first is that Python does not use line terminators like semicolons and tabs or indents are used to indicate new blocks of code instead of braces. Additionally, \texttt{for} and \texttt{if} statements have a different syntax than languages like C and C\texttt{++}. Fortunately, we can work around these issues using the API provided by \texttt{clang::Token} and the \texttt{clang::SourceManager}. Each \texttt{Token} exposes a \texttt{getLocation()} method, which will return information about the token's source location as a \texttt{SourceLocation} instance. The \texttt{SourceManager} provides methods for querying the token \\\texttt{SourceLocation} line and column number. With this information, our token analysis is able to know when a new statement has been encountered (line number) and when we enter a new scope or block of code (column number). 

We provide a \texttt{PyXasmTokenCollector} implementation of \texttt{TokenCollector} which implements the following mapping strategy: (1) decompose the incoming \texttt{clang::Token}s into a list of lines or Python statements and associate each line with its corresponding column number, and (2) for each line, we will parse that line to a corresponding parse tree, walk that tree, and map pertinent nodes to C\texttt{++} equivalents. 
\begin{wrapfigure}{r}{.35\textwidth}
  \lstset {language=C++}
  \begin{lstlisting}
__qpu__ void bell(qreg q) {
    using qcor::pyxasm;
    H(q[0])
    CX(q[0], q[1])
    for i in range(q.size()):
        Measure(q[i])
}
\end{lstlisting}
\caption{A simple \texttt{qcor} C\texttt{++} kernel written in the PyXasm dialect. With the appropriate \texttt{using} statement, \texttt{qcor} can parse this to a valid C\texttt{++} representation using the PyXASM TokenCollector.}
\label{fig:cpp_pyxasm}
\end{wrapfigure}
For this second part, we leverage the Antlr \cite{antlr} parser-generator library using the provided C\texttt{++} runtime. We generate our parser based on the publicly available \texttt{Python3.g4} grammar specification and implement a custom parse tree visitor that will visit certain nodes and perform the mapping to C\texttt{++}.

Overall, our mapping implementation will handle two types of nodes: (1) quantum instructions and (2) classical instructions. The key Python syntax tree node, which we need to handle, is the atom expression node representing call-like expressions. By inspecting the input tokens at this node, we can distinguish if the function call is a quantum intrinsic instruction, a quantum kernel call, or a classical function call.  The first case includes basic quantum gates, such as Hadamard, CNOT gates, etc., which could be identified by matching the name token to the gate registry. The second case is when users make a call to another quantum kernel or their derivatives (controlled or adjoint). There is a special syntax rewrite rule for this case to chain the execution context of this kernel to its sub-kernels, hence requiring a custom syntax handling procedure. 
If no patterns match, then we assume that this a classical function call to utility helpers provided by the QCOR runtime library and keep the function call intact. Control flow nodes, such as \texttt{if} and \texttt{for} statements, are also converted to their C\texttt{++} equivalents. 
\begin{wrapfigure}{r}{.6\textwidth}
  \lstset {language=C++}
  \begin{lstlisting}
class bell : 
    public qcor::QuantumKernel<class bell, qreg> {
protected:
  void operator()(qreg q) {
    ...
    quantum::h(q[0]);
    quantum::cnot(q[0], q[1]);
    for (auto i : range(q.size())) {
      quantum::mz(q[i]);
    }
  }

public:
  bell(qreg q) : QuantumKernel<bell, qreg>(q) {}
  bell(std::shared_ptr<qcor::CompositeInstruction> 
        _parent, qreg q)
      : QuantumKernel<bell, qreg>(_parent, q) {}
  virtual ~bell() {
    ...
    operator()(q);
    ...
  }
  ...
};

void bell(
  std::shared_ptr<qcor::CompositeInstruction> parent, 
  qreg q) {
  class bell __ker__temp__(parent, q);
}
\end{lstlisting}
\caption{Syntax handling result of the PyXASM kernel in Fig.~\ref{fig:cpp_pyxasm}.}
\label{fig:cpp_rewrite}
\end{wrapfigure}

As previously described, we have a stack data structure to keep track of the scope of each code block in order to inject the braces into the rewritten source as appropriate. To rewrite assignment-like statements (e.g., \texttt{a = f(b)}), we keep track of known variables to inject inline variable declarations (via C\texttt{++} \texttt{auto} type keyword) when necessary. This is an example of low-level semantic transformation rules that we implemented to bridge Pythonic imperative expressions to a C\texttt{++} static source. Except for \texttt{Measure} instructions, quantum instructions and kernels don't have return values, thus do not involve in these assignment statements.  

Fig.~\ref{fig:cpp_rewrite} is an example of the rewritten source code generated by the QCOR syntax handler. The Pythonic body of the bell kernel in Fig.~\ref{fig:cpp_pyxasm} is transformed into a call operator of a \texttt{qcor} \texttt{QuantumKernel} subclass. Specifically, quantum instructions are lowered into concrete quantum runtime functions. Similarly, the Python for loop is converted into a valid C\texttt{++} equivalent with proper scoping by braces. 

Kernel invocation is triggered automatically by the destructor, see~\cite{qcor-paper} for a detailed discussion about the QCOR execution model. Essentially, the PyXASM handler has transformed the Pythonic quantum kernel into a native QCOR \texttt{QuantumKernel} subclass, which could be compiled and linked with the QCOR C\texttt{++} runtime.

\section{Quantum Just-in-Time Compilation}
\label{sec:QJIT}
Having a \texttt{TokenCollector} utility that can turn Python-like source strings into valid C\texttt{++} code, we now turn our attention to the just-in-time compilation capability of the \texttt{qcor} compiler. For this purpose, we put forward a \texttt{QJIT} class (Quantum Just-in-Time) which loads a quantum kernel source string containing any \texttt{qcor}-supported dialects and maps it to executable binary code in memory, at runtime. The results of successful \texttt{QJIT} compilation are callable function pointers that can be invoked from C\texttt{++} or Python.   

At the high-level, \texttt{QJIT} provides a \texttt{jit\_compile} method taking as input a source string. This string captures a quantum kernel in any \texttt{qcor}-supported dialects, such as XASM, OpenQASM, or PyXASM. Additionally, we can specify a list of kernel names as dependencies, whose bodies will be pulled in during compilation. This dependency injection mechanism allows us to support nested quantum kernels in the JIT context, whereby kernels are submitted to the \texttt{QJIT} engine individually. Additionally, we allow for extra source code and header include directives to be customized if necessary. The overall execution flow of the \texttt{jit\_compile} procedure is described in Algorithm~\ref{alg:qjit}. When a source string is presented to \texttt{QJIT}, 
\begin{wrapfigure}{r}{0.55\textwidth}
\begin{minipage}{0.55\textwidth}
\begin{algorithm}[H]
\begin{algorithmic}
    \STATE Run Syntax Handler:
    \bindent
        \STATE 1. Analyze the kernel function signature
        \STATE 2. Run \texttt{QCORSyntaxHandler} on the kernel body
        \STATE 3. Assemble the function signature and body.
    \eindent
    \STATE Cache rewritten kernel source (by kernel name)
    \STATE Inject kernel dependency and extra source code (if any)
    \STATE Hash the rewritten kernel source to generate a digest
    \STATE Look up the IR cache with the hash digest
    \IF{CACHE FOUND}
    \bindent
        \STATE Load LLVM IR (bit-code) from cache
    \eindent
    \ELSE
    \bindent
        \STATE Compile the source string and emit LLVM IR
        \STATE Save IR (bit-code) to cache
    \eindent
    \ENDIF
    \STATE Add the LLVM IR to the JIT dynamic library binary
    \STATE Retrieve function pointers of the kernel in the JIT binary
\end{algorithmic}
\caption{QCOR Quantum Just-in-Time compilation workflow.}
\label{alg:qjit}
\end{algorithm}
\end{minipage}
\end{wrapfigure}
it will first execute the syntax handler to translate that invalid quantum source into valid C\texttt{++} code, and then compile that rewritten C\texttt{++} representation to an LLVM IR \texttt{Module}. One key feature of \texttt{QJIT} is its ability to cache the results of both stages for future fast look-up (compile once and reuse the same resultant bitcode). Moreover, the runtime syntax handler results are also cached in memory for future kernel dependency injection. Specifically, as \texttt{QJIT} compiles a kernel, it also saves the rewritten code. Thus, if this kernel is later called within another kernel, the cached source can be injected without any syntax transformation required. Since \texttt{QJIT} works on a kernel-by-kernel basis, this runtime cache eliminates any unnecessary syntax processing when the just-in-time kernels are nested. 

Secondly, the compilation result, in terms of the LLVM IR bit-code, is cached permanently on disk. We use the hash digest of the source as the look-up key to the bit-code file. Therefore, future recompilations of the same kernel can be bypassed by loading the IR from files, as shown in the conditional block of Algorithm 1. This caching mechanism is valuable for the Python use case that we are targeting because the LLVM compilation will only occur once, no matter how many times the Python script with embedded kernels is run. The first time a kernel is seen will result in a longer JIT compilation time, but future invocations of that kernel will be fast due to the runtime loading of cached LLVM bit-code files.

\section{Extending Python via QJIT}
\label{sec:qjit_details}

Equipped with the syntax handling and just-in-time facilities, the qcor compiler infrastructure can be extended to Python. In particular, we want to translate quantum kernels written in the PyXASM dialect along with a subset of Python language features into binary machine code at runtime. Our goal here is to extend the QCOR single-source programming model to Python programmers and thus allow them to explore QCOR or prototype quantum kernels in a familiar setting (e.g.,  in IPython notebooks). In particular, users should be able to utilize common Pythonic language constructs like control flow, e.g., Python if statements and for loops, utility functions, such as range or print, and, more importantly, high-level libraries like \texttt{numpy} and \texttt{openfermion}. Thanks to their linkage to the \texttt{qcor} runtime libraries, the execution of these Pythonic quantum kernels is then seamlessly integrated (just-in-time) with the optimization, placement, and remote job submission pipeline.

Our quantum just-in-time compiler (qjit) in Python operates in three phases: (1) capturing functions intended to describe quantum kernels by means of Python decorators, (2) delegating the compilation of the quantum kernel body to the qcor \texttt{QJIT} compiler, and (3) invoking the compiled functions with runtime parameters provided in Python. In the following, we will describe each of the above tasks in detail.   

\subsection{Python Kernel Decorator - \texttt{qjit}}
Since Python is an interpreted language, it is not compatible, out-of-the-box, with the whole kernel source compilation approach of the \texttt{qcor} compiler. It is also worth noting that implementing quantum runtime imperatively via a binding layer is viable but incurs the performance overhead associated with the Python interpreter. 
\begin{wrapfigure}{r}{.35\textwidth}
  \lstset {language=Python}
  \begin{lstlisting}
@qjit
def bell(q : qreg):
    H(q[0])
    CX(q[0], q[1])
    for i in range(q.size()):
        Measure(q[i])
\end{lstlisting}
\caption{Python kernel expressing the Bell experiment as equivalent to the C\texttt{++} example depicted in Fig.~\ref{fig:cpp_pyxasm}. The \texttt{@} symbol is the standard Python notation to denote that the \texttt{qjit} decorator is to be applied to the function.}
\label{fig:python_bell}
\end{wrapfigure}

Fortunately, Python provides a \emph{decorator} utility, which is part of the language, as a means to apply a transformation to a function. In other words, we can decorate a Python function to redirect the execution of that function to an alternative implementation that we provide. Specifically, in the \texttt{qcor} context, we provide a decorator to record the Python kernel function body and pipe it to the QJIT compiler, as described in the previous section, to create a binary function backed by the \texttt{qcor} runtime. 

Fig.~\ref{fig:python_bell} depicts a decorated quantum kernel whereby \texttt{qcor} provides a \texttt{qjit} decorator implementation activated by the shorthand \texttt{@} symbol. When the interpreter process this function, it will effectively construct a \texttt{qjit} class instance providing the name and signature of the wrapped function to the constructor. This provides an opportunity for \texttt{qcor} to introspect the function body and analyze the function arguments. The qjit class provides an internal \texttt{\_\_call\_\_} method for invocation of the JIT-compiled function which the Python kernel is mapped into.

\subsubsection{Type Inference}
As an dynamically-typed language, Python functions are polymorphic in their inputs, i.e. the number of arguments and their types are flexible. Hence, mapping these functions to \texttt{qcor} kernels, which are strongly-typed C\texttt{++} functions requires some specific constraints to be enforced upon the decorated function. 

First, we require that all function arguments are annotated with type hints. Our Python JIT compiler takes an eager approach, i.e. compiling the kernel at the time of declaration rather than invocation. Therefore, we do not infer the argument types from the specific values provided at the call site. We note that in the vast majority of use cases, \texttt{qcor} users only need to define quantum kernels with fixed argument types. Hence, providing concrete type information enables type-checking, improves kernel declaration readability, prevents any potential type-inference ambiguity, and simplifies our \texttt{qjit} implementation. Second, the Python-to-C\texttt{++} type mapping information must be statically provided. In other words, we provide a map from Python types to their C\texttt{++} equivalents for commonly-used types, including fundamental types (e.g., integer and floating-point numbers), QCOR types (e.g., \texttt{Operator}, \texttt{qreg}, etc.), and containers (list/array) of those types. We use this map to construct the C\texttt{++} function signature compatible with the \texttt{qjit} kernel, thus runtime arguments provided when invoking the decorated function can be unpacked by the JIT-compiled binary.   

We also want to note that \texttt{qjit} kernels support both pass-by-value and pass-by-reference argument types. Since Python does not support passing simple arithmetic types, e.g., integers or floating-point numbers, by reference, we have a custom introspective mechanism to support pass-by-reference for \texttt{qjit} kernel arguments. Using the type annotation data whereby we have defined new reference types that \texttt{qjit} supports, \texttt{qjit} generates the appropriate native (C\texttt{++}) \texttt{QuantumKernel} signature, i.e., deducing the argument type to be a C\texttt{++} reference type (\texttt{Type\&}), see Fig.~\ref{fig:cpp_rewrite}. Having the correct \texttt{QuantumKernel} signature, we turn our attention to the argument reference passing in the Python-C\texttt{++} interop layer. Ultimately, the Pythonic \texttt{qjit} makes copies (pass-by-value) of simple-type arguments when packing runtime variables for invocations. Thus, we need a mechanism to persist these by-reference values across the Python-C\texttt{++} boundary. Fortunately, \texttt{qcor} quantum kernels, by definition, should always take a \texttt{qreg} as the first argument. Taking advantage of the fact that Python objects are always passed-by-reference, we persist any other by-reference variables to the input \texttt{qreg} at the end of the kernel execution by injecting extra code during \texttt{qjit} decoration handling. At invocation time (\texttt{qjit}'s \texttt{\_\_call\_\_}), we introspect the Python interpreter's stack frames to determine the outer scope variables that are passed by reference to the \texttt{qjit} kernel. 
\begin{wrapfigure}{r}{.4\textwidth}
  \lstset {language=Python}
  \begin{lstlisting}
@qjit
def a(q : qreg):
    ....
    # Not using any kernels
@qjit
def b(q : qreg):
    ....
    # Not using any kernels
@qjit
def c(q : qreg):
    ....
    # Call a
    a(q)
@qjit
def d(q : qreg):
    ....
    # Call b and the adjoint of c 
    b(q)
    c.adjoint(q)
\end{lstlisting}
\includegraphics[width=0.3\textwidth]{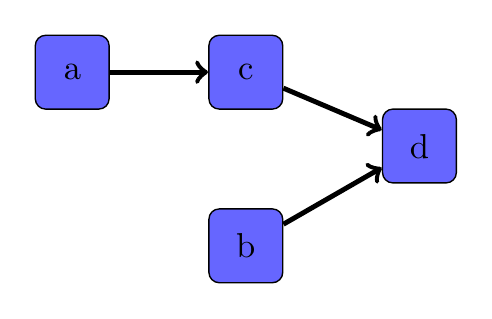}
\caption{Kernel dependency analysis: (top) code snippet demonstrating how one might construct multiple quantum kernels and compose them, (bottom) corresponding directed acyclic graph representation of kernel dependencies in the code snippet.}
\label{fig:qjit_nested}
\end{wrapfigure}
After the JIT kernel execution, we unpack the updated values stored in \texttt{qreg} and assign them to these Python variables accordingly. This passing-by-reference feature is relevant to the FTQC runtime, whereby results of measurement can be accessed in real-time and can be passed around between nested kernels or to the calling code using argument references.

\subsubsection{Closure Analysis}

The \texttt{qcor} \texttt{qjit} decorator is capable of capturing inputs over the scope of the decorated function and passing them to the C\texttt{++} JIT kernel. The common use case is numerical constants that users declared outside the scope of the function, i.e. at the global scope of the Python script. By examining the interpreter stack, we record any global variables that are defined and inject those variables into the rewritten source as local definitions. Another common scenario where lexical capture is required is Python module import aliasing. During the processing of the decorated kernel, we resolve any library import alias names to their original name. For instance, if users alias \texttt{numpy} with a very custom name and use the library via this alias in the kernel body, \texttt{qjit} will be able to resolve it back to the original \texttt{numpy} name before the source is passed to the PyXasm syntax handler and ultimately the C\texttt{++} JIT compiler.      

\subsubsection{Dependency Analysis}
The \texttt{qcor} \texttt{qjit} decorator supports kernel composition allowing users to define quantum kernels in a modular fashion. For example, kernels can call each other so long as there is no circular dependency created. Since we decorate and JIT-compile each kernel separately, the \texttt{qcor} Python library has a mechanism to detect and analyze this dependency information and then provide it to the QJIT's \texttt{jit\_compile} method as explained in the previous section. Fig.~\ref{fig:qjit_nested} (top) is an example of such a scenario whereby there are four kernels, \texttt{a}, \texttt{b}, \texttt{c} and \texttt{d}, that are inter-dependent. The relationship between them can be expressed via a directed acyclic graph (DAG) as shown in Fig.~\ref{fig:qjit_nested} (bottom).

When the \texttt{qjit} constructor is activated and provided with the kernel body, \texttt{qcor} will scan its content to pick up any calls to known kernels which have been presented to the JIT compiler as decorated functions. This scan also covers usage of the controlled or adjoint versions of a kernel which are automatically generated by \texttt{qcor}. For example, kernel \texttt{d} in Fig.~\ref{fig:qjit_nested} makes use of the conjugate of kernel \texttt{c} via the \texttt{adjoint()} method. The relationship between kernels is built up internally as the Python interpreter parses the script and delegates kernel functions to \texttt{qjit}. Using a topological sorting algorithm, we can determine the logical ordering of each kernel dependency to provide to the \texttt{jit\_compile} function. As noted in Sec.~\ref{sec:QJIT}, \texttt{QJIT} will pull the transformed source of these dependencies from the cache when constructing the LLVM JIT module. 

\subsection{Execution Model}
All the processing via Python's decorator utility discussed thus far occurs at the point of kernel definition. The results of this procedure are a decorated Python function (as a Python \texttt{qjit} object) and a binary-compiled function that has been loaded into the memory. Via the binding layer, the \texttt{qjit} object has a handle to the native \texttt{QJIT} instance which kept pointers to those JIT-compiled kernels.

When the kernel is invoked, the interpreter will delegate the invocation to the \texttt{qjit} instance's \texttt{\_\_call\_\_} method and pass along all the runtime arguments provided at the call site. To bridge the gap between the variadic nature of kernel arguments and the statically-defined binding interface of the native \texttt{QJIT} host, we make use of the heterogeneous map utility of XACC, which models the Python \texttt{dict} type, as the argument container. This heterogeneous container stores argument values indexed by their names, which will later be unpacked in a strongly-type manner using the type information provided previously during the JIT compilation procedure.

Kernel execution occurs entirely at the natively-compiled binary level, i.e., no back-and-forth interaction with the Python interpreter. The underlying execution pipeline includes building up the quantum circuit IR tree, applying optimization/placements passes, and submitting the circuit IR to the XACC Accelerator instance (simulator or cloud-based quantum backends). Quantum runtime configurations, such as shot count, backend name, or optimization parameters, can be specified either as command-line arguments (when invoking the whole script with the Python executable) or via API functions that can be used in the script body. Execution results are persisted to the \texttt{qreg} object which is the required first argument for all \texttt{qcor} kernels. 
\begin{wrapfigure}{r}{.55\textwidth}
  \lstset {language=Python}
  \begin{lstlisting}
@qjit
def ccnot(q : qreg):
    ....
    with decompose(q) as ccnot:
        # Create an 8x8 identity matrix
        ccnot = np.eye(8)
        # Modify some entries to create a CCNOT
        ccnot[6,6] = 0.0
        ccnot[7,7] = 0.0
        ccnot[6,7] = 1.0
        ccnot[7,6] = 1.0
    
    # Other instructions, e.g. Measurement
    ....
\end{lstlisting}
\caption{Using the automatic circuit synthesis extension. We overload the standard Python \texttt{with} statement with a special syntax to request circuit synthesis of arbitrary unitary matrices. The \texttt{with} block is translated to a gate sequence at invocation.}
\label{fig:qjit_decompose}
\end{wrapfigure}

As described in~\cite{qcor-paper}, \texttt{qcor} also supports a tightly-coupled execution model, so-called FTQC, allowing for fast feedback of quantum measurement results. \texttt{qjit}-decorated kernels can also be executed on this FTQC runtime. For instance, kernel arguments can be passed by reference in order to pass real-time measurement-controlled values between kernels or between a kernel and host code (pure Python code). One minor feature that we also want to mention is the ability of the syntax handler to translate Python \texttt{print} function to a C\texttt{++} equivalent. Thus, users can inject debug printouts into their Python FTQC kernels and observe the console log when the JIT-compiled binary of them is executed.       

\subsection{Circuit Synthesis Extension}
In~\cite{qcor-paper}, we have introduced the unitary matrix programming capability (\texttt{decompose}) of \texttt{qcor} (C\texttt{++}) whereby users can specify a circuit block in terms of its unitary matrix representation. Similarly, we also expose this unitary decompose functionality to Python \texttt{qjit} as shown in Fig.~\ref{fig:qjit_decompose} where we overload the standard Python \texttt{with} statement to handle a special \texttt{decompose} token. The name of the unitary matrix variable, whose body is constructed in the \texttt{with} statement scoped block, is given after the \texttt{as} keyword. We also want to note that the content of the unitary matrix can be constructed in a native Pythonic way, e.g., via math libraries, such as \texttt{numpy} or \texttt{scipy}~\cite{2020SciPy-NMeth}, or quantum libraries, such as \texttt{cirq}~\cite{cirq}, \texttt{OpenFermion}~\cite{openfermion}, etc.
\begin{wrapfigure}{r}{.5\textwidth}
  \lstset {language=Python}
  \begin{lstlisting}
import numpy as np
@qjit
def ucc1(q : qreg, x : float):
    with compute:
        Rx(q[0], np.pi/2.)
        for i in range(3):
            H(q[i+1])
        for i in range(3):
            CX(q[i], q[i+1])
    with action:
        Rz(q[3], x)

@qjit
def kernel(q: qreg, d : float):
    ucc1.ctrl(q[4], q[0:4], d)

qq = qalloc(5)
kernel.print_kernel(qq, 1.234)
\end{lstlisting}
\caption{Compute-Action syntax in PyXasm and its utility for controlled-circuit generation.}
\label{fig:py_compute_action}
\end{wrapfigure}
Various unitary synthesis techniques are available in \texttt{qcor}, such as \texttt{kak}~\cite{cartan}, \texttt{qfast}~\cite{qfast}, \texttt{qsearch}~\cite{qsearch}, \texttt{qfactor}~\cite{qfactor}, etc. If not specifically given in the \texttt{decompose} call after the qubit register, as the sample in Fig.~\ref{fig:qjit_decompose}, the default method (\texttt{qfast}) will be used. The \texttt{with decompose} code block is completely equivalent to a sub-circuit, i.e., other quantum instructions can be placed before and after this block when constructing the \texttt{qjit} kernel. 

\subsection{Compute-Action Extension}
\begin{figure}[b!]
\centering
\includegraphics[width=0.9\textwidth]{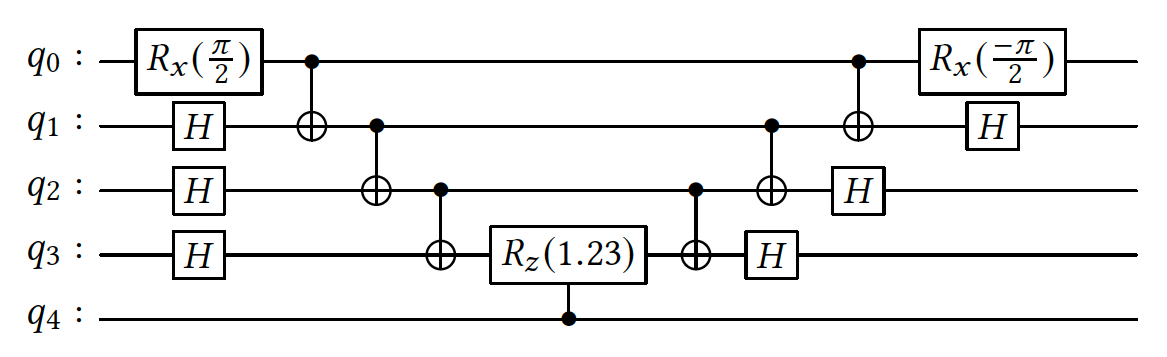}
\caption{Quantum circuit representation of \texttt{kernel} described in Fig.~\ref{fig:py_compute_action}. Although the we apply the controlled version of \texttt{ucc1}, only the \texttt{action} block needs to be transformed ($Rz \mapsto CRz$). The adjoint of the \texttt{compute} block is appended automatically.}
\label{fig:py_compute_action_circ}
\end{figure}
\label{sec:compute-action}
As described in~\ref{sec:qcor_compute_action}, the \emph{compute-action-uncompute} pattern is commonly used to construct quantum computing algorithms. Through the aforementioned extension to the \texttt{qcor} C\texttt{++} runtime we are able to enable this feature in Pythonic \texttt{qjit} kernels in a straightforward manner. Similar to the unitary synthesis syntax, we define special \texttt{with} statements to define the scoped blocks of \texttt{compute} ($U$) and \texttt{action} ($V$). These blocks will be rewritten to the equivalent C\texttt{++} brace-enclosed compute-action syntax. 

Being able to represent this at the programming language level contributes to the improvements of syntax clarity and circuit generation performance. Fig.~\ref{fig:py_compute_action} is an example of using the compute-action-uncompute pattern in a \texttt{qjit} kernel. The compute block ($U$) is scoped within a '\texttt{with compute}' block followed by a '\texttt{with action}' block describing the action unitary ($V$). It is immediately evident that the ability to inject the uncompute sub-circuit ($U^\dagger$) automatically shortens the kernel description and makes the programmers' intent clearer.

More importantly, the compute-action annotation allows for optimal generation of the controlled version of the original kernel. In particular, we only need to apply control on the $V$ unitary, thus significantly simplify the circuit. For instance, the quantum circuit of controlled \texttt{ucc1} in Fig.~\ref{fig:py_compute_action} is shown in Fig.~\ref{fig:py_compute_action_circ} whereby controlling action is only required for the central $Rz$ gate. 

\subsection{Pythonic Kernel Callable - KernelSignature}
\label{sec:py-kernel-signature}
\begin{wrapfigure}{r}{.5\textwidth}
  \lstset {language=Python}
  \begin{lstlisting}
@qjit
def reflect_about_uniform(q: qreg):
    with compute:
        H(q)
        X(q)
    with action:
        Z.ctrl(q[0: q.size() - 1], 
                    q[q.size() - 1])
    
@qjit
def run_grover(q: qreg, 
                oracle_var: 
                    KernelSignature(qreg), 
                iterations: int):
    H(q)
    # Iteratively apply the oracle 
    # then reflect about uniform
    for i in range(iterations):
        oracle_var(q)
        reflect_about_uniform(q)
    # Measure all qubits
    Measure(q)

@qjit
def cz_oracle(q: qreg):
    CZ(q[0], q[2])
    CZ(q[1], q[2])

q = qalloc(3)
run_grover(q, cz_oracle, 1)
\end{lstlisting}
\caption{Passing Python \texttt{qjit} kernels as arguments of type \texttt{KernelSignature}. Specifically, the \texttt{run\_grover} kernel implements a generic Grover's algorithm for arbitrary quantum oracles provided as a callable argument.}
\label{fig:py_grover}
\end{wrapfigure}
As described in~\ref{sec:qcor_QuantumSignature}, the QCOR language allows users to pass quantum functions, i.e., kernels or lambdas, as arguments of type \texttt{KernelSignature} capturing the required invocation signature. In Python, we also expose an equivalent syntax as depicted in Fig.~\ref{fig:py_grover}, whereby a generic Grover's algorithm kernel takes another kernel describing the quantum oracle operation. Invoking the top-level kernel, in this case, involves providing a sub-kernel as its argument. 

Since \texttt{qjit}-wrapped kernels are pure Python objects, they are not compatible with native JIT-compiled kernel calls. In other words, it is illegal to invoke \texttt{run\_grover} with the \texttt{cz\_oracle} \texttt{qjit} object once \texttt{run\_grover} is translated into the JIT function pointer. Technically, what we want to provide as the \texttt{KernelSignature} argument instead is the binary function pointer of \texttt{cz\_oracle}. Thanks to the \texttt{KernelSignature} type annotation, we can reliably perform this transformation by querying the function handle (pointer) of the callable \texttt{qjit} variable provided at the call site (\texttt{\_\_call\_\_} method) and replacing the \texttt{qjit} objects with its native function pointer in the variable pack. 

As shown in Fig.~\ref{fig:py_grover}, to specify a Pythonic \texttt{KernelSignature} type, we provide a variadic list of argument types that the callable kernel expects. This information is necessary for later reconstruction of the underlying templated C\texttt{++} \texttt{KernelSignature} argument from the JIT-compiled function pointer. We also want to note that automatic adjoint and control modifiers are available to callable arguments similar to a kernel calling other kernels from the global scope. Syntactic differences for kernel modifiers, i.e., static class functions or member functions, at the code-generation layer are distinguished based on the scoping information of the variable, e.g., a global kernel instance vs. a variable declared in the local scope.   

\subsection{Decorated \texttt{qjit} Utility Methods}
\label{sec:qjit_utils}
\begin{table}[t]
\caption{List of notable \texttt{qjit} decorator class utility member functions.}
\label{tab:qjit_utils}
\begin{center}
 \begin{tabular}{|p{0.2\linewidth} | p{0.75\linewidth}|} 
 \hline
 Method & Description \\ 
 \hline
 \texttt{extract\_composite} & Convert this \texttt{qjit} quantum kernel into an XACC \texttt{CompositeInstruction}. The returned \texttt{CompositeInstruction} is the final IR tree at the gate-by-gate level retrieved from the QCOR runtime after all internal processing such as optimization and placement. This \texttt{CompositeInstruction} is equivalent to the circuit that would be executed on the selected backend if the kernel is invoked.
  \\ 
 \hline
 \texttt{print\_kernel} & Print the resolved \texttt{CompositeInstruction} of this \texttt{qjit} kernel to the console. \\ 
 \hline
 \texttt{openqasm} & Get the OpenQASM equivalent (as a string) of this \texttt{qjit} kernel. \\ 
 \hline
 \texttt{as\_unitary\_matrix} & Get the unitary matrix representation of this kernel.   \\ 
 \hline
 \texttt{observe} & Construct and execute observable circuits using this kernel as the base ansatz. The observable operator is given to this method along with other arguments that the kernel expects. This method returns the final expectation value after aggregating all the sub-circuit results.  \\ 
 \hline
\end{tabular}
\end{center}
\end{table}
Not only does the \texttt{qjit} decorator class wrapper provide a mechanism to delegate the Python \texttt{\_\_call\_\_} invocation to the JIT-compiled kernel, but it also provides opportunities to add extra functionality to the Pythonic quantum kernel as member functions of the decorated \texttt{qjit} object. In Table~\ref{tab:qjit_utils}, we list a few notable extra utility member functions that we implement for the \texttt{qjit} class. Rather than executing the qjit kernel in a function-call-like manner, users can use these methods to retrieve its internal representation (e.g., as IR tree, unitarty matrix, or OpenQASM string), or compute an observable expectation on the state that the kernel represents.

\section{Demonstration}
\subsection{Variational Quantum Algorithms}
In this example, we use \texttt{qjit} to construct a quantum kernel representing an ansatz for the construction of a \texttt{qcor} \texttt{ObjectiveFunction}. As introduced in~\cite{mintz2019qcor}, \texttt{ObjectiveFunction} is a formal concept in the QCOR specification describing a prototypical $y = F(\textbf{x})$ functional form. In the VQE context, $F(\textbf{x})$ can be the energy function which we want to minimize. Therefore, we need to provide the variational state-preparation circuit (ansatz) as well as the Hamiltonian operator to the \texttt{ObjectiveFunction} generator as shown in Fig.~\ref{fig:demo_vqe}. In this example, the kernel is a custom circuit parameterized by a single variable (\texttt{t0}). The deuteron Hamiltonian (\texttt{H}) is expressed in terms of builtin Pauli spin operators (\texttt{X}, \texttt{Y}, and \texttt{Z}). With the \texttt{ObjectiveFunction} constructed from the JIT-compiled \texttt{ansatz} and the Hamiltonian, we can use any classical optimizers, e.g., \texttt{nlopt}~\cite{nlopt}, \texttt{mlpack}~\cite{mlpack}, \texttt{scipy}~\cite{2020SciPy-NMeth}, etc., to find the ground-state energy via variational minimization. It is worth noting that rather than using the \texttt{Optimizer} and \texttt{ObjectiveFunction} utilities of QCOR to implement the VQE algorithm, one can also use the \texttt{observe} method of the \texttt{ansatz} \texttt{qjit}-decorated object to compute the expectation value (see~\ref{sec:qjit_utils}) as the cost/objective value to integrate with any optimization protocols.

The results of running the script in Fig.~\ref{fig:demo_vqe} are shown in Fig.~\ref{fig:demo_vqe_results}, where we plot the energy values at each optimization iteration using different quantum backends, namely an ideal simulator, a noisy simulator, and a quantum device (\texttt{ibmq\_guadalupe}, 16 qubits, quantum volume of 32). The qcor runtime is retargetable, meaning that the same quantum kernel can be executed on any hardware backends.
\clearpage
\begin{figure}[t!] 
\centering
\begin{subfigure}{.48\textwidth}
  \lstset {language=Python}
  \begin{lstlisting}
from qcor import *
@qjit
def ansatz(q : qreg, t0: float):
  X(q[0])
  Ry(q[1], t0)
  CNOT(q[1],q[0])      

# Define the Hamiltonian
H = -2.1433 * X(0) * X(1) \
    - 2.1433 * Y(0) * Y(1) \
    + .21829 * Z(0) - 6.125 * Z(1) + 5.907

# Create the ObjectiveFunction
obj = createObjectiveFunction(ansatz, H, 1)
# Create the nlopt optimizer and run
optimizer = createOptimizer('nlopt')
results = optimizer.optimize(obj)
\end{lstlisting}
\caption{VQE algorithm with a \texttt{qjit} ansatz}
\label{fig:demo_vqe} 
\end{subfigure}\hspace{12pt}
\begin{subfigure}{.45\textwidth}
\begin{tikzpicture}
\begin{axis}[
      cycle list name=exotic,
      legend columns=3,
      xmin = 0, xmax = 12,
      xlabel = {Optimization Iteration}, 
      ylabel = {Energy Value}, 
      y label style={at={(axis description cs:0.1,0.5)},anchor=south},
      title = {VQE Optimization Results}]
\addplot
table[x=x, y=y] {
x   y
0   -0.43629
1   1.62039983001
2   10.19359983
3   4.45270045348
4   -1.59384659069
5   -1.60946732175
6   -1.18132079196
7   -1.71581973894
8   -1.74876023748
9   -1.70797158239
10  -1.73757413396
11  -1.74743696429
12  -1.74769248655
};
\addlegendentry{sim};
\addplot
table[x=x, y=y] {
x   y
0   -0.229856403809
1   1.9030749585
2   9.48590717041
3   4.1249022876
4   -1.18437001221
5   -1.10394400635
6   -0.859159255371
7   -1.31590825195
8   -1.33309243896
9   -1.27039182861
10  -1.25233861328
11  -1.26704219482
12  -1.25382018555
};
\addlegendentry{sim-noise};
\addplot
table[x=x, y=y] {
x   y
0   -0.303065290527
1   2.04806640625
2   9.25493000732
3   4.17696710449
4   -1.27284730713
5   -1.32037066406
6   -0.820928088379
7   -1.27753792725
8   -1.22089459229
9   -1.35580751221
10  -1.22183096191
11  -1.28020394043
12  -1.27371001465
};
\addlegendentry{\texttt{ibmq\_guadalupe}};
\addplot [domain = 0:12, thick, dashed, red]{-1.74886};
\end{axis}
\end{tikzpicture} 
\caption{Result of executing the code in (a) using an ideal simulator, a noisy simulator, and a physical device (\texttt{ibmq\_guadalupe} 16-qubit backend). The dashed line represents the true ground-state energy.}
\label{fig:demo_vqe_results}
\end{subfigure}
\caption{Variational Quantum Eigensolver using the QCOR specification API and Pythonic quantum kernels annotated with the \texttt{qjit} decorator.}
\label{fig:demo_vqe_all}
\end{figure}
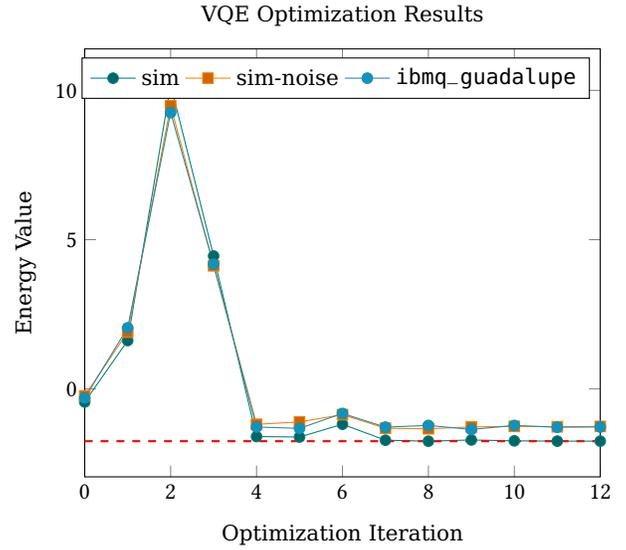
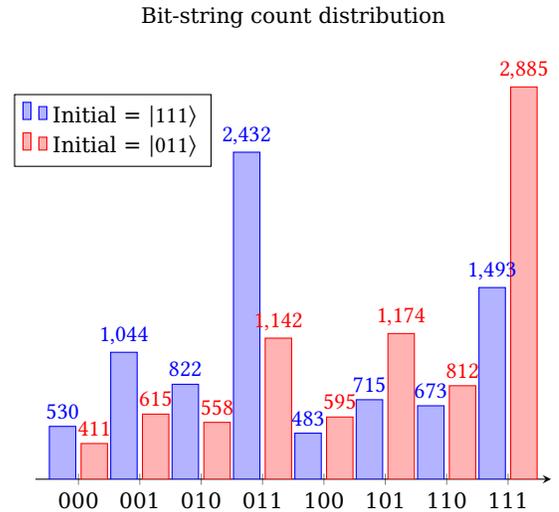
\begin{figure}[b!] 
\centering
\begin{subfigure}{.48\textwidth}
  \lstset {language=Python}
\begin{lstlisting}
from qcor import *
@qjit
def ccnot(q : qreg):
    # create 111
    X(q)
    # decompose ccnot matrix        
    with decompose(q) as ccnot:
        ccnot = np.eye(8)
        ccnot[6,6] = 0.0
        ccnot[7,7] = 0.0
        ccnot[6,7] = 1.0
        ccnot[7,6] = 1.0   
    Measure(q)
# Allocate 3 qubits, run, get result
q = qalloc(3)
ccnot(q)
print(q.counts())
\end{lstlisting}
\caption{Using \texttt{qcor} unitary decompose to synthesize a quantum circuit for a CCNOT gate.}
\end{subfigure}\hspace{12pt}
\begin{subfigure}{.45\textwidth}
\begin{tikzpicture}
\begin{axis}[ybar, title={Bit-string count distribution}, symbolic x coords={000, 001, 010, 011, 100, 101, 110, 111},
  legend pos = north west, axis y line=none, axis x line=bottom, nodes near coords, enlarge x limits=0.1, xtick=data, legend style={at={(0.15,0.9)},anchor=north}] 
\addplot+ coordinates {(000, 530) (001, 1044) (010, 822) (011, 2432) (100, 483) (101, 715) (110, 673) (111, 1493)}; 
\addplot+ coordinates {(000, 411) (001, 615) (010, 558) (011, 1142) (100, 595) (101, 1174) (110, 812) (111, 2885)};
\legend{Initial = $|111\rangle$, Initial = $|011\rangle$}
\end{axis} 
\end{tikzpicture}
\caption{Results of decomposition for two inputs on IBMQ Yorktown. Expected mapping: $|011\rangle \mapsto |111\rangle$ and vice-versa.}
\label{fig:demo_ccnot_synthesis}
\end{subfigure}
\caption{Variational Quantum Eigensolver using the QCOR specification API and Pythonic quantum kernels annotated }
\label{fig:demo_vqe_all}
\end{figure}
\clearpage
Hence, users just need to use the \texttt{-qpu} flag to select the QPU backend that they want to run the quantum experiments on. With an ideal simulator, the observable converges to the expected ground-state energy of deuteron. There is a slight deviation from the theoretical value due to gate noises and readout errors when executing the VQE loop on real hardware. Using a noisy simulator, taking into account the device model, we can reproduce the hardware results with very high fidelity, indicating that the hardware imperfections are well-characterized by the calibration procedure. 

\begin{figure}[b]
  \lstset {language=Python}
  \begin{lstlisting}
from qcor import *
@qjit
def ansatz(q : qreg, x : List[float]):
    X(q[0])
    with decompose(q, kak) as u:
        from scipy.sparse.linalg import expm
        from openfermion.ops import QubitOperator
        from openfermion.transforms import get_sparse_operator
        qop = QubitOperator('X0 Y1') - QubitOperator('Y0 X1')
        qubit_sparse = get_sparse_operator(qop)
        u = expm(0.5j * x[0] * qubit_sparse).todense()

# Define the Hamiltonain
from openfermion.ops import FermionOperator as FOp
H = FOp('', 0.0002899) + FOp('0^ 0', -.43658) + FOp('1 0^', 4.2866) 
    + FOp('1^ 0', -4.2866) + FOp('1^ 1', 12.25) 

# Create the VQE ObjectiveFunction
vqe_obj = createObjectiveFunction(ansatz, H, 1)
# Create an optimizer (gradient free)
optimizer = createOptimizer('nlopt')
# Find the ground state via optimization
results = optimizer.optimize(vqe_obj)
  \end{lstlisting}
  \caption{VQE algorithm where the ansatz circuit is described in terms of a matrix constructed using Scipy and OpenFermion. The Hamiltonian observable is also expressed as FermionOperator.}
\label{fig:demo_integration}
\end{figure}
\subsection{Circuit Synthesis}
In this example, we want to demonstrate the circuit synthesis functionality of \texttt{qcor} whereby a sub-circuit is described by its unitary matrix as shown in Fig.~\ref{fig:demo_ccnot_synthesis} (top). Here, the unitary matrix represents a doubly-controlled NOT gate (CCNOT). We add X gates before the CCNOT block to examine the truth table of the decomposed circuit. For instance, when all qubits are in $|1\rangle$ state, we expect the result to be $|011\rangle$, i.e., the target qubit (MSB) is flipped from $|1\rangle$ to $|0\rangle$. Fig.~\ref{fig:demo_ccnot_synthesis} (bottom) shows the bit-string distribution from executing the circuit (8192 shots) on the IBMQ Yorktown device for the initial states of $|111\rangle$ and $|011\rangle$. We expect transitions from $|011\rangle$ to $|111\rangle$ and vice-versa, which are evident in the distributions in Fig.~\ref{fig:demo_ccnot_synthesis} (bottom). We want to note that this decomposition is using the QFAST \cite{qfast} algorithm and is mainly for 
demonstration purposes since there exists better decompositions specifically for the CCNOT gate. The QFAST-decomposed circuit contains more than 40 CNOT gates\footnote{Avg. CNOT Error:
2.14e-2} which contribute to noisy distribution results.

\subsection{Third-party Integration}
Through this novel Python language extension, we anticipate that \texttt{qcor} users will be able to incorporate a wide range of libraries into their quantum programming workflows. For instance, Fig.~\ref{fig:demo_integration} demonstrates the ability to integrate Open-Fermion~\cite{openfermion} and Scipy~\cite{2020SciPy-NMeth} Python packages in a \texttt{qjit} quantum kernel definition.  

This example essentially runs the same VQE procedure as the one depicted in Fig.~\ref{fig:demo_vqe}. However, rather than defining the Hamiltonian in terms of \texttt{qcor} Pauli operators (e.g., $X$, $Y$, and $Z$), we use the \texttt{FermionOperator} representation provided in the popular \texttt{OpenFermion} Python package. Internally, \texttt{qcor} has compatibility adapters to convert these library-specific objects into \texttt{qcor} representations.

\subsection{Circuit Composition Performance}
\begin{wrapfigure}{r}{.45\textwidth}
  \lstset {language=Python}
  \begin{lstlisting}
@qjit
def trotter_circ(q : qreg, 
        exp_args: List[PauliOperator], 
        n_steps: int):
  for i in range(n_steps):
    for exp_arg in exp_args:
      exp_i_theta(q, 1.0, exp_arg)
\end{lstlisting}
\caption{Composition of Hamiltonian evolution simulation using Trotter decomposition.}
\label{fig:demo_trotter_circ}
\end{wrapfigure}

One of the key design aspects of \texttt{qcor} is its system-level orientation intended for large-scale deployment. Hence, the implementation is entirely based on C\texttt{++} for optimal performance. In~\cite{qcor-paper}, we have demonstrated significant speed-up in terms of quantum circuit composition time between quantum kernels written in C\texttt{++} and Python. 

Since the \texttt{qjit} extension that we presented here is based on JIT compilation into binary code, we expect that the runtime performance of these \texttt{qjit} kernels is on par with their native C\texttt{++} counterparts.  It is worth reiterating that with \texttt{qjit} the whole kernel is wrapped as a binary function handle as described in Sec.~\ref{sec:qjit_details}, hence skipping all the expensive interactions with the Python interpreter during kernel execution.  

\begin{table}[h!]
\caption{Trotter circuit composition performance (not including PySCF time). Number of Trotter steps = 1. Hamiltonian from PySCF using sto3g basis. We skip the Qiskit run for C$_2$H$_4$ due to time limit.}
\centering
 \begin{tabular}{|c | c | c | c | c | c |} 
 \hline
 Molecule & Qubits & Terms & Instructions & qjit runtime [sec] & Qiskit runtime [sec] \\ 
 \hline
 H$_2$ & 4 & 14 & 82 & 0.00465 & 0.1569 \\ 
 H$_2$O & 14 & 1085 & 17963 & 0.2441 & 22.9048 \\ 
 N$_2$ & 20 & 2950 & 62618 & 0.7128  & 82.6061 \\ 
 C$_2$H$_4$ & 28 & 57092 & 1503664 & 16.84 & -- \\ 
 \hline
\end{tabular}
\label{tab:perf_comp}
\end{table}

To evaluate the circuit composition performance of the \texttt{qjit} kernel, we time the \texttt{trotter\_circ} kernel runtime, as defined in Fig.~\ref{fig:demo_trotter_circ}, against a set of molecular Hamiltonian operators, as shown in Table~\ref{tab:perf_comp}. In this benchmark, we use a fixed number of Trotter steps (one step) and feed the kernel with the Pauli-sum representation\footnote{via Jordan-Wigner transformation} of each molecule's Hamiltonian (generated by PySCF~\cite{sun2018pyscf}). 

The timing data in Table~\ref{tab:perf_comp} only include the duration of the circuit construction, i.e., from the time the kernel function is invoked to the point when the flattened instruction list is generated. This includes looping over individual terms, generating and appending the \texttt{exp\_i\_theta} sub-circuits. On average, our JIT-compiled kernel function is 30x-100x faster in constructing the flattened gate IR than a pure Python equivalent. The main contribution to the performance improvement is that, for \texttt{qjit} kernels, all PyXASM constructs, 
such as for loops, \texttt{exp\_i\_theta} instructions, are compiled to native C\texttt{++} binary objects resulting in a highly optimal and efficient execution. This provides a high-performance and scalable quantum programming environment, not only for near-term use cases but also for future large-scale deployments.   

\subsection{FTQC Runtime Execution}
\begin{wrapfigure}{r}{.45\textwidth}
  \lstset {language=Python}
  \begin{lstlisting}
@qjit
def applyQEC(q : qreg):
    ancIdx = 3
    # q[0] and q[1] parity
    CX(q[0], q[ancIdx])
    CX(q[1], q[ancIdx])
    parity01 = Measure(q[ancIdx])
    #Reset anc qubit
    Reset(q[ancIdx])
    # q[1] and q[2] parity
    CX(q[1], q[ancIdx])
    CX(q[2], q[ancIdx])
    parity12 = Measure(q[ancIdx])
    #Reset anc qubit
    Reset(q[ancIdx])
    # Error decode and correction
    # Compute the error syndrome value:
    parity = 0
    if parity01:
        parity = parity + 1
    if parity12:
        parity = parity + 2
    print("Syndrome value=", parity)
    # Apply correction
    if parity == 1: 
        X(q[0])
    if parity == 2:
        X(q[2])
    if parity == 3:
        X(q[1])
\end{lstlisting}
  \caption{Apply a quantum-error-correction round on a logical qubit encoded in the three-qubit bit-flip code using FTQC runtime. The console logging via \texttt{print()} and syndrome calculation reflects the real-time feedback of measurement results available in the FTQC runtime.}
\label{fig:demo_ftqc}
\end{wrapfigure}
In~\cite{qcor-paper}, we introduced a fully dynamic quantum runtime for \texttt{qcor}, named FTQC, capable of handling flexible control flow based on real-time measurement results. This runtime is relevant to quantum error correction whereby syndrome measurement results are used to detect and correct quantum errors. Fig.~\ref{fig:demo_ftqc} illustrates the FTQC runtime utility in simulating quantum error correction code. In this case, it is a simple three-qubit repetition code toy model which could detect single bit-flip (X) errors by comparing the parities between qubit pairs. The code snippet in Fig.~\ref{fig:demo_ftqc} is a \texttt{qjit} kernel subroutine that projects the parity of neighboring qubits (\texttt{q[0]}-\texttt{q[1]} and \texttt{q[1]}-\texttt{q[2]}) into an ancilla qubit. Since the measurement result of this ancilla qubit is immediately available in the FTQC runtime, users can perform arithmetic operations based on bit results as well as conditional gate instructions. 
Using a simple mapping from parity syndrome to error location, we can then correct a bit-flip error if it has occurred. Despite its simplicity, this example demonstrates the versatility of the \texttt{qcor} programming framework. With the introduction of the OpenQASM version 3, we anticipate that this dynamic quantum programming model will soon become available from hardware providers, and through this JIT extension, Python users will be able to take full advantage of the FTQC runtime to develop and test algorithms on tightly-coupled CPU-QPU machine models.

\subsection{Integration with HPC Simulators}
The \texttt{qcor} runtime, based on the XACC framework, incorporates a couple of HPC quantum circuit simulators capable of distributing the compute workload across multiple nodes via the Message Passing Interface (MPI). For instance, XACC has a tensor network-based simulator, named TNQVM~\cite{nguyen2021tensor}, capable of running on leadership-class supercomputers, such as the Summit supercomputer\footnote{200 petaFLOPS, the second fastest supercomputer in the world (as of April 2021)} at Oak Ridge National Laboratory. In this demonstration, we seek to demonstrate the usage of such high-performance simulators to simulate large-scale quantum circuits expressed as \texttt{qjit} kernels.

Fig.~\ref{fig:demo_hpc} is a code snippet whereby we simulate the seminal random circuit sampling circuit that Google ran on their Sycamore chip (53 qubits)~\cite{Arute_2019}. In this example, we use the \texttt{observe} utility of \texttt{qjit}, see Table~\ref{tab:qjit_utils}, to compute the expectation of an arbitrary operator. For demonstration purposes, we compute the all-$Z$ ($Z_0Z_1...Z_{52}$) expectation value. This Python script can be executed in an MPI-distributed environment (e.g. via \texttt{mpiexec}) using packages such as MPI for Python (\texttt{mpi4py})~\cite{mpi4py} as shown in the running command in Fig.~\ref{fig:demo_hpc}. Another feature that we want to highlight in this example is the ability to pass detailed accelerator configurations in a file to the backend accelerator via the \texttt{qcor\_qpu\_config} key. This functionality is often required by complex \texttt{Accelerator} implementations, which have customized performance switches, e.g., memory buffer size, task distribution strategies, etc.  
\begin{figure}[t!] 
\centering
\begin{subfigure}{.48\textwidth}
  \lstset {language=Python}
\begin{lstlisting}
from qcor import *
@qjit
def rcs_circuit(q : qreg):
  ...
  Rz(q[0], -0.78539816339)
  ...
  Rz(q[52], -1.2305673521313445)
  ...    
  fSim(q[47], q[51], 
   1.4908807480931237, 0.48862437201319)
  fSim(q[50], q[52], 
   1.6162569997269376, 0.5014289362839901)
  ...

# Allocate 53 qubits
q = qalloc(53)
# Compute the expectation value
obs = rcs_circuit.observe(ham, q)
<@\textcolor{red}{------------------- Run with ---------------------}@>
$ mpiexec -n <n_procs> \
 python3 sycamore.py \
 -qpu tnqvm[qcor_qpu_config:<config_file>]
\end{lstlisting}
\caption{Simulating \texttt{qjit} kernel on HPC simulator (TNQVM). The quantum kernel is a random circuit sampling (RCS) adapting from the Sycamore quantum supremacy experiment involving 53 qubits.}
\label{fig:demo_hpc}
\end{subfigure}\hspace{12pt}
\begin{subfigure}{.45\textwidth}
\begin{tikzpicture}
\pgfplotsset{%
    width=\textwidth,
    height=0.8\textwidth
}
\begin{axis}[
    ybar,
    tick label style={font=\small},
    label style={font=\small},
    xticklabels={1, 2, 4, 8}, 
    xtick={1,2,3,4},
    ymin=0,
    xlabel={Number of MPI processes},
    ylabel={Elapsed Time [sec]},
    y label style={at={(axis description cs:0.15,.5)},anchor=south},
    ]
    \addplot +[bar shift=-.0cm, area legend] coordinates {(1, 253.819774) (2, 169.938532) (3, 89.684513)};
\end{axis}
\begin{axis}[
    axis y line*=right,
    y label style={at={(axis description cs:1.45,.5)},anchor=south},
    axis x line=none,
    ylabel = {GFlops/sec},
    scaled y ticks = false,
    ytick style={draw=none},
    ymin=0.1, ymax=1.5,
    ]
    \addplot +[bar shift=.0cm, area legend, style={mark=*}] coordinates {(1, 0.355) (2, 1.31) (3, 1.36)};
\end{axis}
\end{tikzpicture}
\caption{Total simulation runtime (bar, left axis) and tensor processing Flop rate (line, right axis)  vs. number of MPI processes (\texttt{n\_procs})\footnotemark of running the code sample in Fig.~\ref{fig:demo_hpc} with the TNQVM simulator. The total runtime also includes non-distributed tasks such as tensor network optimization. The flop count is computed based on the actual number of numerical operations performed.}
\label{fig:perf_results}
\end{subfigure}
\end{figure}

\footnotetext{Compute node: 2x AMD EPYC 7302 16-core Processor (3.0 GHz)}

A sample performance scaling data is shown in Fig.~\ref{fig:perf_results}, where we vary the number of MPI processes to assess the task distribution. We also want to note that this demonstration only uses CPU's hence not utilizing the full capability of the TNQVM simulator, which could perform tensor processing on GPU's thanks to its integration with the ExaTN library~\cite{exatn}. As we can see from Fig.~\ref{fig:perf_results}, there is a significant performance boost in terms of Flop rate of the ExaTN numerical backend when going from one to two MPI processes, which matches our expectation for this two-socket workstation. The performance saturates when going to higher MPI ranks due to machine configuration. We also want to note that the total runtime includes not only the tensor contraction task, whose Flop count we track, but also other workloads, such as tensor network construction and contraction sequence optimization. For example, the improvement in total runtime in the rank-4 case is mostly due to faster contraction optimization and better contraction path. 

\section{Conclusion}
We have presented a language extension to Python for heterogeneous quantum-classical computing that utilizes quantum kernel just-in-time compilation at the C\texttt{++} level to ensure a performant execution model. Our work builds upon the \texttt{qcor} C\texttt{++} infrastructure to enable a single-source programming model for quantum computing that is quantum coprocessor retargetable. Our work seeks to provide a future-proof platform for rapid prototyping and experimentation of quantum scientific computing applications under both loosely and tightly coupled CPU-QPU machine models. We anticipate that this core Pythonic infrastructure can serve as a foundation for rapid experimentation via scripting, application libraries, and benchmarking of quantum-classical use-cases in a hardware-agnostic manner. 

\section*{Acknowledgment}
This work has been supported by the US Department of Energy (DOE) Office of Science Advanced Scientific Computing Research (ASCR) Accelerated Research in Quantum Computing (ARQC) and Quantum Computing Application Teams (FWP No. ERKJ347). AIDE-QC supported the development of the LLVM JIT for quantum kernel processing. QCAT supported the implementation of the QCOR specification for the Python language. This research used resources of the Oak Ridge Leadership Computing Facility, which is a DOE Office of Science User Facility supported under Contract DE-AC05-00OR22725. ORNL is managed by UT-Battelle, LLC, for the US Department of Energy under contract no. DE-AC05-00OR22725.

\bibliographystyle{plain}
\bibliography{main}
\end{document}